\input amssym.tex
\input epsf
\epsfclipon


\magnification=\magstephalf
\hsize=14.0 true cm
\vsize=19 true cm
\hoffset=1.0 true cm
\voffset=2.0 true cm

\abovedisplayskip=12pt plus 3pt minus 3pt
\belowdisplayskip=12pt plus 3pt minus 3pt
\parindent=1.0em


\font\sixrm=cmr6
\font\eightrm=cmr8
\font\ninerm=cmr9

\font\sixi=cmmi6
\font\eighti=cmmi8
\font\ninei=cmmi9

\font\sixsy=cmsy6
\font\eightsy=cmsy8
\font\ninesy=cmsy9

\font\sixbf=cmbx6
\font\eightbf=cmbx8
\font\ninebf=cmbx9

\font\eightit=cmti8
\font\nineit=cmti9

\font\eightsl=cmsl8
\font\ninesl=cmsl9

\font\sixss=cmss8 at 8 true pt
\font\sevenss=cmss9 at 9 true pt
\font\eightss=cmss8
\font\niness=cmss9
\font\tenss=cmss10

 at 12 true pt
 at 12 true pt
\font\bigrm=cmr10 at 12 true pt
 at 12 true pt
 at 12 true pt

\font\Bigi=cmmi12 at 16 true pt
 at 16 true pt
\font\Bigrm=cmr12 at 16 true pt
 at 16 true pt
 at 16 true pt

\catcode`@=11
\newfam\ssfam

\def\tenpoint{\def\rm{\fam0\tenrm}%
    \textfont0=\tenrm \scriptfont0=\sevenrm \scriptscriptfont0=\fiverm
    \textfont1=\teni  \scriptfont1=\seveni  \scriptscriptfont1=\fivei
    \textfont2=\tensy \scriptfont2=\sevensy \scriptscriptfont2=\fivesy
    \textfont3=\tenex \scriptfont3=\tenex   \scriptscriptfont3=\tenex
    \textfont\itfam=\tenit                  \def\it{\fam\itfam\tenit}%
    \textfont\slfam=\tensl                  \def\sl{\fam\slfam\tensl}%
    \textfont\bffam=\tenbf \scriptfont\bffam=\sevenbf
    \scriptscriptfont\bffam=\fivebf
                                            \def\bf{\fam\bffam\tenbf}%
    \textfont\ssfam=\tenss \scriptfont\ssfam=\sevenss
    \scriptscriptfont\ssfam=\sevenss
                                            \def\ss{\fam\ssfam\tenss}%
    \normalbaselineskip=13pt
    \setbox\strutbox=\hbox{\vrule height8.5pt depth3.5pt width0pt}%
    \let\big=\tenbig
    \normalbaselines\rm}

\def\ninepoint{\def\rm{\fam0\ninerm}%
    \textfont0=\ninerm      \scriptfont0=\sixrm
                            \scriptscriptfont0=\fiverm
    \textfont1=\ninei       \scriptfont1=\sixi
                            \scriptscriptfont1=\fivei
    \textfont2=\ninesy      \scriptfont2=\sixsy
                            \scriptscriptfont2=\fivesy
    \textfont3=\tenex       \scriptfont3=\tenex
                            \scriptscriptfont3=\tenex
    \textfont\itfam=\nineit \def\it{\fam\itfam\nineit}%
    \textfont\slfam=\ninesl \def\sl{\fam\slfam\ninesl}%
    \textfont\bffam=\ninebf \scriptfont\bffam=\sixbf
                            \scriptscriptfont\bffam=\fivebf
                            \def\bf{\fam\bffam\ninebf}%
    \textfont\ssfam=\niness \scriptfont\ssfam=\sixss
                            \scriptscriptfont\ssfam=\sixss
                            \def\ss{\fam\ssfam\niness}%
    \normalbaselineskip=12pt
    \setbox\strutbox=\hbox{\vrule height8.0pt depth3.0pt width0pt}%
    \let\big=\ninebig
    \normalbaselines\rm}

\def\eightpoint{\def\rm{\fam0\eightrm}%
    \textfont0=\eightrm      \scriptfont0=\sixrm
                             \scriptscriptfont0=\fiverm
    \textfont1=\eighti       \scriptfont1=\sixi
                             \scriptscriptfont1=\fivei
    \textfont2=\eightsy      \scriptfont2=\sixsy
                             \scriptscriptfont2=\fivesy
    \textfont3=\tenex        \scriptfont3=\tenex
                             \scriptscriptfont3=\tenex
    \textfont\itfam=\eightit \def\it{\fam\itfam\eightit}%
    \textfont\slfam=\eightsl \def\sl{\fam\slfam\eightsl}%
    \textfont\bffam=\eightbf \scriptfont\bffam=\sixbf
                             \scriptscriptfont\bffam=\fivebf
                             \def\bf{\fam\bffam\eightbf}%
    \textfont\ssfam=\eightss \scriptfont\ssfam=\sixss
                             \scriptscriptfont\ssfam=\sixss
                             \def\ss{\fam\ssfam\eightss}%
    \normalbaselineskip=10pt
    \setbox\strutbox=\hbox{\vrule height7.0pt depth2.0pt width0pt}%
    \let\big=\eightbig
    \normalbaselines\rm}

\def\tenbig#1{{\hbox{$\left#1\vbox to8.5pt{}\right.\n@space$}}}
\def\ninebig#1{{\hbox{$\textfont0=\tenrm\textfont2=\tensy
                       \left#1\vbox to7.25pt{}\right.\n@space$}}}
\def\eightbig#1{{\hbox{$\textfont0=\ninerm\textfont2=\ninesy
                       \left#1\vbox to6.5pt{}\right.\n@space$}}}

\font\sectionfont=cmbx10
\font\subsectionfont=cmti10

\def\figurecaptionfont{\ninepoint}
\def\tablecaptionfont{\ninepoint}
\def\footnotefont{\eightpoint}


\newcount\equationno
\newcount\bibitemno
\newcount\figureno
\newcount\tableno

\equationno=0
\bibitemno=0
\figureno=0
\tableno=0


\footline={\ifnum\pageno=0{\hfil}\else
{\hss\rm\the\pageno\hss}\fi}


\def\section #1. #2 \par
{\vskip0pt plus .10\vsize\penalty-100 \vskip0pt plus-.10\vsize
\vskip 1.6 true cm plus 0.2 true cm minus 0.2 true cm
\global\def\equationlabel{#1}
\global\equationno=0
\leftline{\sectionfont #1. #2}\par
\immediate\write\terminal{Section #1. #2}
\vskip 0.7 true cm plus 0.1 true cm minus 0.1 true cm
\noindent}


\def\subsection #1 \par
{\vskip0pt plus 0.8 true cm\penalty-50 \vskip0pt plus-0.8 true cm
\vskip2.5ex plus 0.1ex minus 0.1ex
\leftline{\subsectionfont #1}\par
\immediate\write\terminal{Subsection #1}
\vskip1.0ex plus 0.1ex minus 0.1ex
\noindent}


\def\appendix #1. #2 \par
{\vskip0pt plus .20\vsize\penalty-100 \vskip0pt plus-.20\vsize
\vskip 1.6 true cm plus 0.2 true cm minus 0.2 true cm
\global\def\equationlabel{\hbox{\rm#1}}
\global\equationno=0
\leftline{\sectionfont Appendix #1. #2}\par
\immediate\write\terminal{Appendix #1. #2}
\vskip 0.7 true cm plus 0.1 true cm minus 0.1 true cm
\noindent}



\def\equation#1{$$\displaylines{\qquad #1}$$}
\def\enum{\global\advance\equationno by 1
\hfill\llap{{\rm(\equationlabel.\the\equationno)}}}
\def\noenum{\hfill}
\def\next#1{\cr\noalign{\vskip#1}\qquad}


\def\ifundefined#1{\expandafter\ifx\csname#1\endcsname\relax}

\def\ref#1{\ifundefined{#1}?\immediate\write\terminal{unknown reference
on page \the\pageno}\else\csname#1\endcsname\fi}

\newwrite\terminal
\newwrite\bibitemlist

\def\bibitem#1#2\par{\global\advance\bibitemno by 1
\immediate\write\bibitemlist{\string\def
\expandafter\string\csname#1\endcsname
{\the\bibitemno}}
\item{[\the\bibitemno]}#2\par}

\def\beginbibliography{
\vskip0pt plus .15\vsize\penalty-100 \vskip0pt plus-.15\vsize
\vskip 1.2 true cm plus 0.2 true cm minus 0.2 true cm
\leftline{\sectionfont References}\par
\immediate\write\terminal{References}
\immediate\openout\bibitemlist=biblist
\frenchspacing\parindent=1.8em
\vskip 0.5 true cm plus 0.1 true cm minus 0.1 true cm}

\def\endbibliography{
\immediate\closeout\bibitemlist
\nonfrenchspacing\parindent=1.0em}


\def\figurecaption#1{\global\advance\figureno by 1
\narrower\figurecaptionfont
Fig.~\the\figureno. #1}

\def\tablecaption#1{\global\advance\tableno by 1
\vbox to 0.5 true cm { }
\centerline{\tablecaptionfont%
Table~\the\tableno. #1}
\vskip-0.4 true cm}

\def\thicktablerule{\hrule height1pt}
\def\thintablerule{\hrule height0.4pt}

\tenpoint

\immediate\openin\bibitemlist=biblist
\ifeof\bibitemlist\immediate\closein\bibitemlist
\else\immediate\closein\bibitemlist
\input biblist \fi


\def\thismonth{\ifcase\month\or
January\or February\or March\or April\or May\or June\or
July\or August\or September\or October\or November\or December\fi}



\def\rmd{{\rm d}}

\def\rme{{\rm e}}
\def\rmO{{\rm O}}


\def\rz{{\Bbb R}}
\def\gz{{\Bbb Z}}
\def\nz{{\Bbb N}}

\def\Re{{\rm Re}\,}


\def\proof{\noindent{\sl Proof:}\kern0.6em}

\def\frac#1#2{\hbox{$#1\over#2$}}
\def\dual{\mathstrut^*\kern-0.1em}

\def\lvec#1{\setbox0=\hbox{$#1$}
    \setbox1=\hbox{$\scriptstyle\leftarrow$}
    #1\kern-\wd0\smash{
    \raise\ht0\hbox{$\raise1pt\hbox{$\scriptstyle\leftarrow$}$}}
    \kern-\wd1\kern\wd0}
\def\rvec#1{\setbox0=\hbox{$#1$}
    \setbox1=\hbox{$\scriptstyle\rightarrow$}
    #1\kern-\wd0\smash{
    \raise\ht0\hbox{$\raise1pt\hbox{$\scriptstyle\rightarrow$}$}}
    \kern-\wd1\kern\wd0}
\def\slash#1{\setbox0=\hbox{$#1$}\setbox1=\hbox{$\kern1pt/$}
    #1\kern-\wd0\kern1pt/\kern-\wd1\kern\wd0}


\def\nabstar#1{{\nabla\kern0.5pt\smash{\raise 4.5pt\hbox{$\ast$}}
               \kern-5.5pt_{#1}}}
\def\drv#1{{\partial_{#1}}}
\def\drvstar#1{{\partial\kern0.5pt\smash{\raise 4.5pt\hbox{$\ast$}}
               \kern-6.0pt_{#1}}}
\def\ldrv#1{{\lvec{\,\partial}_{#1}}}
\def\ldrvstar#1{{\lvec{\,\partial}\kern-0.5pt\smash{\raise 4.5pt\hbox{$\ast$}}
               \kern-5.0pt_{#1}}}


\def\fm{{\rm fm}}




\def\diracstar#1#2{
    \setbox0=\hbox{$\gamma$}\setbox1=\hbox{$\gamma_{#1}$}
    \gamma_{#1}\kern-\wd1\kern\wd0
    \smash{\raise4.5pt\hbox{$\scriptstyle#2$}}}


\def\SUn{{\rm SU}(N)}

\def\Ad{{\rm Ad}\kern0.1em}


\def\Z{{\cal Z}}


\def\R{{\frak R}}
\def\sm{{\frak s}}

%
\rightline{CERN-PH-TH/2004-110}
\rightline{MPP-2004-67}

\vskip 1.0cm 
\centerline{\Bigrm String excitation energies in SU(\kern-1pt{\Bigi N}\/) 
                   gauge theories}
\vskip 0.3 true cm
\centerline{\Bigrm beyond the free-string approximation}
\vskip 0.6 true cm
\centerline{\bigrm Martin L\"uscher}
\vskip1ex
\centerline{\it CERN, Physics Department, TH Division}
\centerline{\it CH-1211 Geneva 23, Switzerland}
\vskip 0.4 true cm
\centerline{\bigrm Peter Weisz}
\vskip1ex
\centerline{\it Max-Planck-Institut f\"ur Physik}
\centerline{\it D-80805 Munich, Germany}
\vskip 0.8 true cm
\thintablerule
\vskip 2.0ex
\ninepoint
\leftline{\bf Abstract}
\vskip 1.0ex\noindent
In the presence of a static quark--antiquark pair,
the spectrum of the low-lying states in SU($N$) gauge theories 
is discrete and likely to be
described, at large quark
separations $r$, by an effective string theory.
The expansion of the excitation energies
in powers of $1/r$, which derives from the latter,
involves an increasing
number of unknown couplings that characterize the 
string self-interactions. 
Using open--closed string duality, we show that
the possible values of the
couplings are constrained by a set of algebraic relations. 
In particular, the corrections of order $1/r^2$ must vanish,
while the $1/r^3$ terms (which we work out for the few lowest
levels) depend on a single adjustable coupling only.

\vskip 2.0ex
\thintablerule

\tenpoint

\vskip-0.2cm

\section 1. Introduction

The formation of string-like flux tubes in SU($N$) gauge theories
is not a proved fact, but the results obtained in lattice gauge
theory over the years leave little doubt
that this physical picture is basically correct.
Recent high-precision numerical simulations
now even confirm the presence of characteristic 
string fluctuation effects in the static quark potential 
[\ref{LWstring},\ref{Pushan}]
and in ratios of loop correlation functions
[\ref{CaselleEtAl}--\ref{CaselleHasenbuschPaneroII}].

A puzzling aspect of these findings is, however, that
string behaviour appears to set in 
at quark--antiquark separations $r$ as small as $0.5$ 
to $1\,\fm$ or so,
where the energy distribution in the ground state of the gauge field
is still far from being tube-like. 
Moreover, at these distances the spectrum of 
the low-lying excited states is poorly matched by the 
regular pattern (with level splittings equal to $\pi/r$) 
that is obtained in the case of a free string with 
fixed ends [\ref{MichaelPerantonis},\ref{KutiEtAlI},\ref{Pushan}].
Closed flux tubes wrapping around a compact space dimension have
also been considered, with similar conclusions
[\ref{ParisiPetronzioRapuano}--\ref{KutiEtAlII}].

At present none of these results exclude that 
the observed mismatches
will gradually disappear at larger distances $r$.
The energy spectrum should in any case be compared
with the predictions of 
the effective string theory of
refs.~[\ref{Nambu}--\ref{UniversalTerm}], in which the
string self-interactions are taken into account in a 
systematic manner.
According to this theory,
the free-string energies are only
the leading terms in an asymptotic expansion in powers of $1/r$,
while the higher-order corrections arise from interaction terms 
of increasing dimensions.
There are finitely many terms at each order 
and a corresponding number of effective coupling constants.

Our principal goal in this paper is to work out the expansion of the 
few lowest energy levels up to order $1/r^3$.
We shall
first show, however, that 
the a priori unknown values of the couplings in the effective theory 
must satisfy certain algebraic relations in order to be consistent with
the closed-string interpretation of the 
partition function.
As a consequence, corrections proportional to $1/r^2$ are excluded
and the number of independent couplings at the next order
of the large-$r$ expansion
is reduced from $2$ to $1$.
The predictive power of the 
effective theory is thus significantly enhanced.

\section 2. Effective string theory

In this section we define the effective string theory
and explain which properties of the gauge theory
it is expected to describe. 
We also compute the string partition function to next-to-leading order,
partly for illustration and partly because this will serve
as a starting point for the discussion in the following section.
Much of what is being said here
already appeared in ref.~[\ref{LWstring}], 
and we refer the reader
to this paper for a less abbreviated presentation.

\subsection 2.1 Definition

Eventually we are interested in the string energy spectrum
in $4$ space-time dimensions, but string formation 
is also studied in $3$ dimensions and we therefore
consider the effective theory in any dimension $d\geq3$.
The fluctuations of a string with fixed ends
can be parametrized by a displacement vector $h(z)$,
with $d-2$ components in the directions orthogonal to the 
string axis, where
\equation{
  z=(z_0,z_1), \qquad 0\leq z_1\leq r,
  \enum
} 
are the world-sheet coordinates. 
Fixed ends means
that the field satisfies 
\equation{
  \left.h(z)\right|_{z_1=0}=\left.h(z)\right|_{z_1=r}=0
  \enum
}
for all times $z_0$, and in the time direction
we shall impose periodic boundary conditions, with period $T$,
at least as long as we
are concerned with the string partition function.

The action of the effective theory is written in the form of 
a series
\equation{
   S=\sigma rT+\mu T+S_0+S_1+\cdots,
   \qquad
   S_0=\frac{1}{2}\int\rmd^D\kern-1pt z
   \left(\partial_ah\partial_ah\right),
   \enum
}
in which $\sigma$ denotes the string tension, $\mu$ an arbitrary
mass parameter and $S_1,S_2,\ldots$ string self-interaction terms of
increasing dimension. For the calculation of the 
string energy levels it is convenient to choose 
the metrics in space-time and on the world-sheet
to be euclidean. 
We shall use dimensional regularization and thus
extend the world-sheet to $D=2-\epsilon$ dimensions
at this point (see appendix A).

As is generally the case in
field theories on manifolds with boundaries
[\ref{QFTbI},\ref{QFTbII}],
the possible interaction terms are localized either on the world-sheet or 
at its boundaries. They may only depend
on the derivatives of $h(z)$ 
and must respect the obvious symmetries of the system. Moreover, 
any terms
that are formally removable by a field transformation can
be dropped, since
their effects on the energy spectrum amount to a renormalization of
the coupling constants multiplying the interactions of lower dimension.

Inspection now shows that 
\equation{
   S_1=\frac{1}{4}b\int\rmd^{D-1}\kern-1pt z
   \left\{
   \left(\partial_1h\partial_1h\right)_{z_1=0}+
   \left(\partial_1h\partial_1h\right)_{z_1=r}
   \right\}
   \enum
}
is the only possible interaction term with coupling $b$ of 
dimension $[\hbox{length}]$.
At second order the couplings have dimension $[\hbox{length}]^2$
and the complete list of terms is
\equation{
   S_2=\frac{1}{4}c_2M^{\epsilon}\int\rmd^D\kern-1pt z
   \left(\partial_ah\partial_ah\right)\left(\partial_bh\partial_bh\right),
   \enum
   \next{2ex}
   S_3=\frac{1}{4}c_3M^{\epsilon}\int\rmd^D\kern-1pt z
   \left(\partial_ah\partial_bh\right)\left(\partial_ah\partial_bh\right),
   \enum
}
where $M$ denotes the renormalization scale and the brackets indicate 
in which way the space indices are contracted.
Since we are using
dimensional regularization, there are no power divergences 
and $b$ is thus not renormalized. In principle an additive renormalization
of $c_2,c_3$ proportional to $b^2$ 
may be required, but it turns out that there are no logarithmic
divergences to this order.

\subsection 2.2 Partition function and the Polyakov loop correlation function

In the effective theory the partition function and the
correlation functions of the field $h(z)$
can be expanded in powers of the couplings $b,c_2,c_3,\ldots$
following the usual steps. From this point of view the theory
looks just like any other non-renormalizable
quantum field theory. In particular, the perturbation series
really is a low-energy expansion, where the higher-order terms
are proportional to inverse powers 
of the external distances such as $T$ and $r$.

To leading order the exact expression for 
the partition function 
\equation{
  \Z_0=
  \rme^{-\sigma rT-\mu T}
  \eta(q)^{2-d},
  \enum
  \next{2ex}
  \eta(q)=q^{1\over24}\prod_{n=1}^{\infty}(1-q^n),
  \qquad
  q=\rme^{-\pi T/r},
  \enum
}
is known since a long time
(see refs.~[\ref{DietzFilk},\ref{AmbjornEtAl},\ref{CaselleEtAl}] for example;
the index ``0" reminds us that we are dealing with the free-string theory).
When expanded in powers of $q$, this formula leads to the
representation
\equation{
  \Z_0=\sum_{n=0}^{\infty}w_n\rme^{-E_n^0T},
  \enum
}
with some positive integral weights $w_n$ and energies $E_n^0$ given by
\equation{
  E_n^{0}=\sigma r+\mu+{\pi\over r}\left\{-\frac{1}{24}(d-2)+n\right\}.
  \enum
}
In particular, the splitting between successive 
energy levels is equal to $\pi/r$ to this order of 
the perturbation expansion.

In the presence of a static quark--antiquark pair, the spectrum of the
low-lying states in the (pure) $\SUn$ gauge theory is surely more
complicated than this, but it is conceivable that the effective string
theory provides an asymptotically correct description of the energy
levels at large distances $r$.
The partition function in this sector of the gauge theory 
is proportional to the Polyakov loop correlation function 
$\left\langle P(x)^{\ast}\kern-1pt P(y)\right\rangle$
at distance $r=|\vec{\kern0.5pt x\kern-0.5pt}-\vec{\kern0.5pt y\kern-0.5pt}|$
on a euclidean space-time cylinder with circumference $T$
in the time direction [\ref{LWstring}].
Evidently, if the energy spectra match (including multiplicities),
the same must be true for the associated partition functions.

We would like to emphasize, however,
that the (full) string partition function $\Z$ 
is only expected to match the Polyakov loop correlation function
to any finite order of the perturbation expansion
in the effective theory, i.e.~only at large $T$ and $r$
and up to higher-order terms.
Decays of highly excited states through glueball radiation
are in fact not included in the string theory,
at least not obviously so,
and this must be reflected by 
relative differences in the partition functions 
of order $\rme^{-mT}$, where $m$ denotes the mass of the lightest
glueball.

\subsection 2.3 Partition function at next-to-leading order

The order 
at which a given interaction term must first be included
in the perturbation expansion
coincides with the dimension of the associated coupling constant.
At next-to-leading order the partition function thus reads
\equation{
  \Z=\Z_0\left\{1-\left\langle S_1\right\rangle_0\right\},
  \enum
}
where the expectation value $\langle\ldots\rangle_0$ is to be evaluated
in the free-string theory. If we introduce the free propagator,
\equation{
  \left\langle h_k(z)h_l(w)\right\rangle_0=\delta_{kl}G(z,w),
  \enum
}
the first-order correction term assumes the form
\equation{
  \left\langle S_1\right\rangle_0=
  \frac{1}{4}b\,(d-2)\int\rmd^{D-1}\kern-1pt z\,\bigl\{
  \left.\drv{1}G(z,z)\ldrv{1}\right|_{z_1=0}+
  \left.\drv{1}G(z,z)\ldrv{1}\right|_{z_1=r}\bigr\},
  \enum
}
and the rules of dimensional regularization must now be invoked
to work it out.

Deferring the details of the calculation to appendix A,
we note that there are no ultra-violet divergences
and quote the result
\equation{
  \left\langle S_1\right\rangle_0=
  -b\left(d-2\right){\partial\over\partial r}\ln\eta(q).
  \enum
}
The correction thus amounts
to a shift $r\to r-b$ in the second 
factor of the leading-order expression (2.7)
(this simple rule actually extends to the next order in $b$, but we 
shall not need to know this here).

\section 3. Open--closed string duality and its consequences

We now proceed to show that the values of the coupling constants 
$b,c_2,c_3,\ldots$ in the effective string theory 
are constrained by certain algebraic relations if the 
Polyakov loop correlation function is to be matched by the
string partition function at large $T$ 
and $r$.

\subsection 3.1 Dual interpretation of the Polyakov loop correlation function

Usually the Polyakov lines are considered to run along the time axis,
forming closed loops through the periodic boundary conditions
in this direction.
This leads to the interpretation 
of the Polyakov loop correlation function as 
the partition function of the gauge theory in the presence
of a static quark--antiquark pair.

We may, however, 
just as well take the compact dimension to be space-like,
in which case the field-theoretical interpretation of the 
Polyakov loop $P(x)$ is that it creates (or annihilates) a closed flux tube
wrapping around the world at, say, time $x_1$ and transverse spatial 
position $x_{\perp}=(x_2,\ldots,x_{d-1})$. 
The loop correlation function at zero transverse momentum 
is therefore expected to have a spectral representation of the form
[\ref{TransferI}--\ref{TransferIV}]
\equation{
  \int\rmd x_{\perp}\left\langle P(x)^{\ast}\kern-1pt P(y)\right\rangle
  =\sum_{n=0}^{\infty}\left|v_n\right|^2\rme^{-\tilde{E}_n\left|x_1-y_1\right|},
  \enum
}
where the exponents $\tilde{E}_n$ and the coefficients $v_n$
are the
energies and transition matrix elements of the possible intermediate
states\kern1.5pt\footnote{$\dagger$}{\footnotefont%
As in the case of the flux tubes with fixed ends,
the higher excited states are unstable against glueball radiation
and the spectrum of these states is thus continuous.
This subtlety
can be ignored in the following, since we shall 
be exclusively interested in the low energy values.}. 
As shown in appendix B,
this implies that the correlation function itself
can be expanded in a series of Bessel functions according to
\equation{
  \left\langle P(x)^{\ast}\kern-1pt P(y)\right\rangle
  =\sum_{n=0}^{\infty}\left|v_n\right|^2
  2r\left({\tilde{E}_n\over2\pi r}\right)^{{1\over2}(d-1)}
  K_{{1\over2}(d-3)}(\tilde{E}_nr).
  \enum
}
While the energy values and the transition matrix elements are not known,
this equation (and similarly the spectral representation 
in the open-string channel) severely constrain the possible analytic form of 
the Polyakov loop correlation function.

\subsection 3.2 Open--closed string duality at leading order 

An interesting question is now whether
the partition function in the effective string theory 
can be expanded, at large $T$ and $r$,
in a series of Bessel functions of the form (3.2).
If not this would imply that the effective theory 
cannot possibly provide an asymptotically
exact description of the Polyakov loop correlation function.

Using the modular transformation property
of the $\eta$-function,
\equation{
  \eta(q)=\left({2r\over T}\right)^{1\over2}\eta(\tilde{q}),
  \qquad
  \tilde{q}=\rme^{-4\pi r/T},
  \enum
}
the leading-order expression (2.7) for the partition function
can be written as a series of exponentials,
\equation{
  \Z_0=
  \rme^{-\mu T}\left({T\over2r}\right)^{{1\over2}(d-2)}
  \sum_{n=0}^{\infty}w_n\rme^{-\tilde{E}_n^0r},
  \enum
}
where the weights $w_n$ are the same as in the expansion (2.9)
of the partition function in the open-string channel.
Evidently, the closed-string energies
\equation{
  \tilde{E}_n^0=\sigma T+{4\pi\over T}
  \left\{-\frac{1}{24}(d-2)+n\right\}
  \enum
}
must coincide with 
the energies $\tilde{E}_n$ in eq.~(3.2),
up to terms of order $b/T^2$, if the series
(3.2) and (3.4) are to match at large $T$ and $r$.

Since their arguments
grow proportionally to $\sigma Tr$ in this limit,
the Bessel functions in the series (3.2)
may be expanded,
\equation{
  \left\langle P(x)^{\ast}\kern-1pt P(y)\right\rangle
  =\sum_{n=0}^{\infty}\left|v_n\right|^2
  \left({\tilde{E}_n\over2\pi r}\right)^{{1\over2}(d-2)}
  \rme^{-\tilde{E}_nr}
  \left\{1+{(d-2)(d-4)\over8\tilde{E}_nr}+\cdots\right\},
  \enum
}
where, to leading order, we may 
replace $\tilde{E}_n$ by $\tilde{E}_n^0$ and
drop all subleading
terms in the curly bracket.
The free-string partition function (3.4) is then seen to 
be of the required form, in any dimension $d$,
and open--closed string duality thus holds to this order
of the perturbation expansion.

\subsection 3.3 Duality at next-to-leading order

The only interaction term that contributes
at next-to-leading order 
is the boundary term $S_1$, and 
the partition function to this order is given by
\equation{
  \Z=\left(1-\sigma Tb\right)\Z_0-b
  {\partial\over\partial r}\Z_0
  \noenum
  \next{2.5ex}
  {\phantom{\Z}}=
  \rme^{-\mu T}\left({T\over2r}\right)^{{1\over2}(d-2)}
  \sum_{n=0}^{\infty}w_n
  \left\{1+b\kern1pt(\tilde{E}_n^0-\sigma T)+{b\over2r}(d-2)\right\}
  \rme^{-\tilde{E}_n^0r}
  \enum
}  
(cf.~subsection 2.3).
From the discussion of the leading-order expression of the par\-ti\-tion
function we know that $\tilde{E}_n=\tilde{E}_n^0+\rmO(b)$,
and once this relation is inserted in eq.~(3.6) it becomes clear that
the $b/r$ term in the curly bracket in eq.~(3.7) cannot be matched.
We thus conclude that
\equation{
  b=0
  \enum
}
is the only possible value of this coupling, which is consistent with
the requirement of open--closed string duality.

\subsection 3.4 Duality at higher orders

At the next order of perturbation theory
there are two string interaction terms, $S_2$ and $S_3$, with 
couplings $c_2$ and $c_3$. Their contribution to 
the partition function is discussed in section 4, where we 
shall find that open--closed string duality holds
if and only if
\equation{
  \left(d-2\right)c_2+c_3={d-4\over2\sigma}.
  \enum
}
The number of adjustable couplings is thus reduced to $1$,
and actually to $0$
in dimension $d=3$, since
the interactions $S_2$ and $S_3$ are proportional to each other 
in this case (cf.~section~6.2).

It is quite clear that the requirement of duality
will lead to similar constraints at any order of perturbation theory.
While this cannot be rigorously excluded, it seems unlikely 
that too many equations
will be obtained at some point, because there are exact
string partition functions with the correct properties
in both the open and the closed string channels.
The formal canonical quantization of the Nambu--Goto string, for example,
yields such a partition function
(appendix C).

\section 4. Partition function at two-loop order

We now always set $b=0$ so that the first non-trivial
corrections to the free-string partition function come
from the quartic interactions $S_2$ and $S_3$. 
At two-loop order the partition function is then given by 
\equation{
   \Z=\Z_0\left\{1-\langle S_2\rangle_0-\langle S_3\rangle_0\right\},
   \enum
}
where, as before, $\Z_0$ and the brackets $\langle\ldots\rangle_0$
stand for the partition function and the 
expectation values in the free theory.

\subsection 4.1 Computation of %
                $\langle S_2\rangle_0$ and $\langle S_3\rangle_0$ 

The contraction of the fields in the interaction terms
$S_2$ and $S_3$ results in tadpole diagrams with two
loops attached to the interaction vertex. We evaluated them
using dimensional regularization, as described in appendix A,
but the same diagrams were actually already calculated many years ago by 
Dietz and Filk [\ref{DietzFilk}] who employed a $\zeta$-function finite-part
prescription.

In terms of the series
\equation{
   \sm_1(q)=q{\partial\over\partial q}\ln\eta(q)=
   {1\over24}-\sum_{n=1}^{\infty}nq^n\left(1-q^n\right)^{-1},
   \enum
   \next{2ex}
   \sm_2(q)=-q{\partial\over\partial q}\sm_1(q)=
   \sum_{n=1}^{\infty}n^2q^n\left(1-q^n\right)^{-2}
   \enum
}
(which are closely related to the first and second Eisenstein series),
the result of the computation reads
\equation{
   \langle S_2\rangle_0=c_2\left(d-2\right){\pi^2T\over2r^3}
   \left\{\left(d-1\right)\sm_2(q)+2\sm_1(q)^2\right\},
   \enum
   \next{2ex}
   \langle S_3\rangle_0=c_3\left(d-2\right){\pi^2T\over2r^3}
   \left\{\frac{1}{2}\left(d+1\right)\sm_2(q)+
   \left(d-1\right)\sm_1(q)^2\right\}.
   \enum
}
In particular, the partition function
is again a power series in $q=\rme^{-\pi T/r}$, as in the case 
of the free-string theory.

\subsection 4.2 Energy spectrum

Once the string interactions are switched on, the accidental
degeneracies
of the energy eigenstates, which are characteristic of the free-string
theory, are not guaranteed to persist. 
The interpretation of the 
partition function as a spectral sum must take this possibility 
into account, and we thus write
\equation{
  \Z=\sum_{n=0}^{\infty}\sum_{i=1}^{k_n}w_{n,i}\kern1pt\rme^{-E_{n,i}T},
  \enum
}
where $k_n$ is the number of different energy values $E_{n,i}$
that are equal to $E_n^0$ to lowest order. 
Evidently the associated multiplicities $w_{n,i}$
must satisfy the sum rule
\equation{
   \sum_{i=1}^{k_n}w_{n,i}=w_n
   \enum
}
for all $n\geq0$.

It follows from this discussion that
the partition function at two-loop order is of the general form
\equation{
  \Z=\sum_{n=0}^{\infty}w_n\left\{1-[E_n^1]\kern1pt T\right\}
  \rme^{-E_n^0T},
  \enum
}
where $[E_n^1]$ is a linear combination of $c_2$ and $c_3$.
This is just the weighted average, with weights 
equal to $w_{n,i}/w_n$, of the first-order energy shifts
$E_{n,i}^1$
in the multiplet with principal quantum number $n$. 
In particular,
\equation{
  [E_0^1]=\left({\pi\over24}\right)^2{d-2\over2r^3}
  \left\{2c_2+\left(d-1\right)c_3\right\},
  \enum
  \next{2ex}
  [E_1^1]=[E_0^1]+{\pi^2\over24r^3}
  \left\{\left(12d-14\right)c_2+\left(5d+7\right)c_3\right\},
  \enum
}
and the average energy shifts of the higher levels can be obtained
straightforwardly by
expanding the partition function in powers of $q$.
Note that the degeneracies of
the ground state and of the first excited states are unchanged
by the interactions.
There is thus a single level in these two cases, with 
first-order energy shift $E_{n,1}^1=[E_n^1]$.

Using the fact that the series $\sm_1$ and $\sm_2$ are
derivatives of the free-string partition function,
the result
\equation{
  [E_n^1]=[E_0^1]+{\pi^2\over r^3}\kern1pt
  n\left[\frac{1}{12}\left(d-2\right)-n\right]c_2
  \quad\hbox{if}\quad c_3=-2c_2,
  \enum
}
is now also easily established.
Couplings related in this way are a
rather special case in many respects. 
In particular, the techniques introduced
in section~5 can be used to show that 
the free-string degeneracies of the energy levels are 
preserved by the interactions at this order of 
perturbation theory.

\subsection 4.3 Closed-string representation

For simplicity we set $\mu=0$ in the following, since
the associated exponential factor only complicates the formulae,
but is of no importance for the 
open--closed string duality of the partition function.
The duality property of the free-string partition function
may then be compactly expressed through
\equation{
  \Z_0=\left(T\over2r\right)^{{1\over2}(d-2)}
  \left.\Z_0\vphantom{q}\right|_{T\leftrightarrow 2r},
  \enum
}
where the notation on the right implies 
a substitution of $T$ by $2r$ and simultaneously of $r$ by $T/2$.

Similar identities can be derived for the first-order perturbations 
(4.4) and (4.5) since these are obtained from the $\eta$-function
by differentiation. Explicitly one finds that
\equation{
  \langle S_2+S_3\rangle_0=
  \left.\langle S_2+S_3\rangle_0\vphantom{q}\right|_{T\leftrightarrow 2r}  
  \noenum
  \next{2ex}
  \phantom{\langle S_2+S_3\rangle_0=\,}
  +\left[\left(d-2\right)c_2+c_3\right]
  \left(d-2\right){4\pi\over T^2}\left[\sm_1(\tilde{q})-{T\over16\pi r}
  \right],
  \enum
}
and the closed-string representation of the partition function thus reads
\equation{
  \Z=\left(T\over2r\right)^{{1\over2}(d-2)}\,
  \biggl\{
  \left.\Z\vphantom{q}\right|_{T\leftrightarrow 2r}
  \noenum
  \next{2ex}
  \phantom{\Z=\,}
  +\left[\left(d-2\right)c_2+c_3\right]
  \left[{d-2\over4Tr}-\sigma-{1\over T}{\partial\over\partial r}\right]
  \left.\Z_0\vphantom{q}\right|_{T\leftrightarrow 2r}
  \biggr\}.
  \enum
}
When expanded in powers of $\tilde{q}$, this formula
yields the series
\equation{
  \Z=\sum_{n=0}^{\infty}
  w_n\left(T\over2r\right)^{{1\over2}(d-2)}\rme^{-\tilde{E}_n^0r}\,
  \biggl\{1-[\tilde{E}_n^1]\kern1pt r
  \noenum
  \next{2ex}
  \phantom{\Z=\,}
  +\left[\left(d-2\right)c_2+c_3\right]
  \left[{4\pi\over T^2}\left\{-\frac{1}{24}\left(d-2\right)+n\right\}
  +{d-2\over4Tr}\right]\biggr\},
  \enum
}
where
the average closed-string energy shifts are given by
\equation{
  [\tilde{E}_n^1]=2\left.[E_n^1]\vphantom{q}\right._{r\to T/2}.
  \enum
}
As in the case of the open-string channel, 
it is not possible, however, to deduce the 
degeneracy pattern of the energy levels from these equations.

The comparison with the right-hand side of eq.~(3.6) now
shows that open--closed string duality holds to this order
of perturbation theory if and only if
the couplings $c_2$ and $c_3$ are related to each other as in eq.~(3.9).
In $d=4$ dimensions this implies, incidentally, that 
the couplings lie on the special line $c_3=-2c_2$, where the 
accidental degeneracies of the energy levels
in the open-string channel are not lifted by the interactions.

\section 5. Computation of the energy shifts $E_{n,i}^1$

The first-order shifts
of the individual energy levels can be extracted from 
the connected correlation functions of the field $h(z)$
at large time separations (we set $T=\infty$ in this section).
Exactly which
correlation functions are considered should not matter
in principle, but the computations can be simplified by introducing
fields that create and annihilate definite free-string modes.
We now first explain this and then proceed with 
a discussion of the symmetry classification of the low-lying
energy eigenstates and 
of the associated eigenvalue equation.

\subsection 5.1 Creation and annihilation operators

Let us consider the fields
\equation{
  a^{\pm}_k(z_0,p)=\int_0^r\rmd z_1\int_0^L\rmd z_2\ldots\rmd z_{D-1}\,
  \sin(pz_1)\left(p\pm\partial_0\right)h_k(z),
  \enum
}
where $p=n\pi/r$, $n\in\nz$, is 
an arbitrary spatial momentum and $k=1,\ldots,d-2$
the space index that labels the directions transverse
to the string axis.
Since we intend to use dimensional regularization, 
the world-sheet is extended to $D$ dimensions, as before, and
an integration over the extra coordinates is included in eq.~(5.1).
It is straightforward to show that
\equation{
  \left\langle  a^{+}_k(z_0,p)h_l(w)\right\rangle_0
  =\delta_{kl}\theta(w_0-z_0)\kern1pt\rme^{-p(w_0-z_0)}\sin(pw_1),
  \enum
  \next{2ex}
  \left\langle  a^{-}_k(z_0,p)h_l(w)\right\rangle_0
  =\delta_{kl}\theta(z_0-w_0)\kern1pt\rme^{-p(z_0-w_0)}\sin(pw_1),
  \enum
}
which suggests that $a^{+}_k(z_0,p)$ and $a^{-}_k(z_0,p)$
create and annihilate harmonic string modes at time $z_0$ 
with wave number $n=pr/\pi$.
The normalization of the associated one-mode states 
can be inferred from
\equation{
  \left\langle  a^{-}_k(z_0,p)a^{+}_l(w_0,q)\right\rangle_0
  =\delta_{kl}\delta_{pq}\theta(z_0-w_0)\kern1pt\rme^{-p(z_0-w_0)}
  rpL^{-\epsilon},
  \enum
}
and we also note that
\equation{
  \left\langle  a^{+}_k(z_0,p)a^{+}_l(w_0,q)\right\rangle_0
  =
  \left\langle  a^{-}_k(z_0,p)a^{-}_l(w_0,q)\right\rangle_0
  =0,
  \enum
}
which is entirely in line with the interpretation of 
$a^{+}_k(z_0,p)$ and $a^{-}_k(z_0,p)$ as creation and annihilation operators
(for simplicity equal-time contact terms were
dropped in these equations).

Products of these fields create and annihilate states with 
any number of excited string modes. Effectively we have the 
association
\equation{
   a^{+}_{k_1}(z_0,p_1)\ldots a^{+}_{k_m}(z_0,p_m)
   \quad\longleftrightarrow\quad
   |n_1,\ldots,n_m;k_1,\ldots,k_m\rangle
   \enum
}
between products of creation operators and the states in the
free-string Fock space with 
transverse space-time indices $k_1,\ldots,k_m$
and wave numbers 
\equation{
   n_j={p_jr\over\pi}, \quad j=1,\ldots,m.
   \enum
}
Wick's rule ensures, incidentally, that  
the scalar products of these states are correctly given by
the correlation functions of the corresponding products of 
the $a^{+}$ and $a^{-}$ fields (the fields
should be inserted at times infinitesimally different from
each other so that contact terms do not arise).

\subsection 5.2 Symmetry considerations

The Fock states (5.6) transform as tensors of rank $m$ under
the group $\rmO(d-2)$ of rotations about the string axis.
Since the effective string theory is rotation-invariant, the 
energy eigenstates can be classified according to 
the irreducible representations of this group.

\topinsert
\newdimen\digitwidth
\setbox0=\hbox{\rm 0}
\digitwidth=\wd0
\catcode`@=\active
\def@{\kern\digitwidth}
\tablecaption{Energy eigenstates with quantum number $n\leq3$} 
\vskip-1.0ex
$$\vbox{\settabs\+&%
                  xxxxxx&xx&
                  xxxxxxxxxxxxxxx&xx&
                  ixxxxxxxxxxxxxxxxxxx&xxxx&
                  xxxxxxxxxxxxxxxxxxxx&\cr
\thicktablerule
\vskip1.0ex
                \+& \hfill$n,i$\hfill
                 && \hfill{\tenrm State}\hfill
                 && \hfill{\tenrm $\rmO(d-2)$ multiplet}\hfill
                 && \hfill$w_{n,i}$\hfill
                 &\cr
\vskip1.0ex
\thintablerule
\vskip1.5ex
  \+& \hfill $0,1$\hfill
  &&  \hskip1ex$|0\rangle$\hfill 
  &&  \hfill{\tenrm scalar}\hfill
  &&  \hskip1ex$1$\hfill
  &\cr
\vskip0.3ex
  \+& \hfill $1,1$\hfill
  &&  \hskip1ex$|1;k\rangle$\hfill 
  &&  \hfill{\tenrm vector}\hfill
  &&  \hskip1ex$d-2$\hfill
  &\cr
\vskip0.3ex
  \+& \hfill $2,1$\hfill
  &&  \hskip1ex$|1,1\rangle$\hfill 
  &&  \hfill{\tenrm scalar}\hfill
  &&  \hskip1ex$1$\hfill
  &\cr
\vskip0.3ex
  \+& \hfill $2,2$\hfill
  &&  \hskip1ex$|2;k\rangle$\hfill 
  &&  \hfill{\tenrm vector}\hfill
  &&  \hskip1ex$d-2$\hfill
  &\cr
\vskip0.3ex
  \+& \hfill $2,3$\hfill
  &&  \hskip1ex$|1,1;\{k,l\}\rangle$\hfill 
  &&  \hfill{\tenrm symmetric $2$-tensor}\hfill
  &&  \hskip1ex$\frac{1}{2}d(d-3)$\hfill
  &\cr
\vskip0.3ex
  \+& \hfill $3,1$\hfill
  &&  \hskip1ex$|1,2\rangle$\hfill 
  &&  \hfill{\tenrm scalar}\hfill
  &&  \hskip1ex$1$\hfill
  &\cr
\vskip0.3ex
  \+& \hfill $3,2$\hfill
  &&  \hskip1ex$|3;k\rangle$\hfill 
  &&  \hfill{\tenrm vector}\hfill
  &&  \hskip1ex$d-2$\hfill
  &\cr
\vskip0.3ex
  \+& \hfill $3,3$\hfill
  &&  \hskip1ex$|1,1,1;k\rangle$\hfill 
  &&  \hfill{\tenrm vector}\hfill
  &&  \hskip1ex$d-2$\hfill
  &\cr
\vskip0.3ex
  \+& \hfill $3,4$\hfill
  &&  \hskip1ex$|1,2;\{k,l\}\rangle$\hfill 
  &&  \hfill{\tenrm symmetric $2$-tensor}\hfill
  &&  \hskip1ex$\frac{1}{2}d(d-3)$\hfill
  &\cr
\vskip0.3ex
  \+& \hfill $3,5$\hfill
  &&  \hskip1ex$|1,2;[k,l]\rangle$\hfill 
  &&  \hfill{\tenrm antisymmetric $2$-tensor}\hfill
  &&  \hskip1ex$\frac{1}{2}(d-2)(d-3)$\hfill
  &\cr
\vskip0.3ex
  \+& \hfill $3,6$\hfill
  &&  \hskip1ex$|1,1,1;\{k,l,j\}\rangle$\hfill 
  &&  \hfill{\tenrm symmetric $3$-tensor}\hfill
  &&  \hskip1ex$\frac{1}{6}(d-2)(d+2)(d-3)$\hfill
  &\cr
\vskip1.0ex
\thicktablerule
}
$$
\vskip-2ex
\endinsert

For general dimension $d$ and principal quantum number 
$n=n_1+\cdots+n_m\leq3$,
the Fock space decomposes into altogether $11$ multiplets, as shown in table~1.
The states listed in the table are
Fock states, or linear combinations of these,
that transform according to some irreducible
representation of $\rmO(d-2)$. As usual the ground state is
denoted by $|0\rangle$ and the states that are non-trivial 
combinations of Fock states are given by
\equation{
  |1,1\rangle=\sum_{k=1}^{d-2}|1,1;k,k\rangle,
  \enum
  \next{2ex}
  |1,1;\{k,l\}\rangle=|1,1;k,l\rangle-{\delta_{kl}\over d-2}\,|1,1\rangle,
  \enum
  \next{2ex}
  |1,2\rangle=\sum_{k=1}^{d-2}|1,2;k,k\rangle,
  \enum
  \next{2ex}
  |1,2;\{k,l\}\rangle={1\over2}\left\{|1,2;k,l\rangle+|1,2;l,k\rangle\right\}
  -{\delta_{kl}\over d-2}\,|1,2\rangle,
  \enum
  \next{2ex}
  |1,2;[k,l]\rangle={1\over2}\left\{|1,2;k,l\rangle-|1,2;l,k\rangle\right\},
  \enum
  \next{2ex}
  |1,1,1;k\rangle=\sum_{l=1}^{d-2}|1,1,1;k,l,l\rangle,
  \enum
  \next{2ex}
  |1,1,1;\{k,l,j\}\rangle=|1,1,1;k,l,j\rangle
  \noenum
  \next{1ex}
  \phantom{|1,1,1;\{k,l,j\}\rangle=\,}
  -{\delta_{kl}\over d}\,|1,1,1;j\rangle
  -{\delta_{jk}\over d}\,|1,1,1;l\rangle
  -{\delta_{lj}\over d}\,|1,1,1;k\rangle.
  \enum
}
It follows from this classification of states that they
will not mix with each other when the 
interactions are switched on, except perhaps for
the two vector multiplets at $n=3$ that are indistinguishable
from the symmetry point of view.

\subsection 5.3 Eigenvalue equation

For specified quantum number $n$ and
representation of $\rmO(d-2)$, 
the states listed in table~1 
span a subspace of lowest-order energy eigenstates
with definite symmetry properties.
The corresponding first-order energy shifts $E_{n,i}^1$
are obtained from the matrix elements of the interaction
\equation{
  S_{\rm int}=S_2+S_3
  \enum
}
in these subspaces.

Let $\phi^{+}_{A}(z_0)$ and $\phi^{-}_{A}(z_0)$
be the expressions 
in creation and annihilation operators at time $z_0$
that are associated to the basis vectors of a given subspace
at level $n$.
We label them by an arbitrary index $A$
and define the normalization matrix $N_{AB}$ through
\equation{
   \left\langle
   \phi^{-}_{A}(t)\phi^{+}_{B}(0)
   \right\rangle_0=N_{AB}\kern1pt\rme^{-n\pi t/r}
   \enum
}
(we take $t$ positive).
The interaction matrix $H_{AB}$ is similarly 
given by the correlation function
\equation{
   \left\langle
   \phi^{-}_{A}(t)S_{\rm int}\phi^{+}_{B}(0)
   \right\rangle_0=
   \left\{H_{AB}t+\cdots\right\}
   \rme^{-n\pi t/r},
   \enum
}   
where the ellipsis stands for any terms independent of $t$.

From the spectral representation of the correlation functions
in the interacting theory, it is now straightforward to 
deduce that the solutions $\lambda$ of 
the generalized eigenvalue problem
\equation{
  Hv=\lambda Nv
  \enum
}
coincide with the first-order excitation energies
\equation{
  \Delta E_{n,i}^1=E_{n,i}^1-E_{0,1}^1.
  \enum
}
If the representation of $\rmO(d-2)$ in the chosen subspace is
irreducible, the eigenvalue problem is trivial, i.e.~$H$ is proportional 
to $N$ and there is a single eigenvalue $\lambda$.
The $n=3$ vector multiplets are an exceptional case from this point of view, 
where one has two eigenvalues
that correspond to $\Delta E_{3,2}^1$ and $\Delta E_{3,3}^1$.

\topinsert
\vbox{
\vskip0.0cm
\epsfxsize=7.0cm\hskip2.5cm\epsfbox{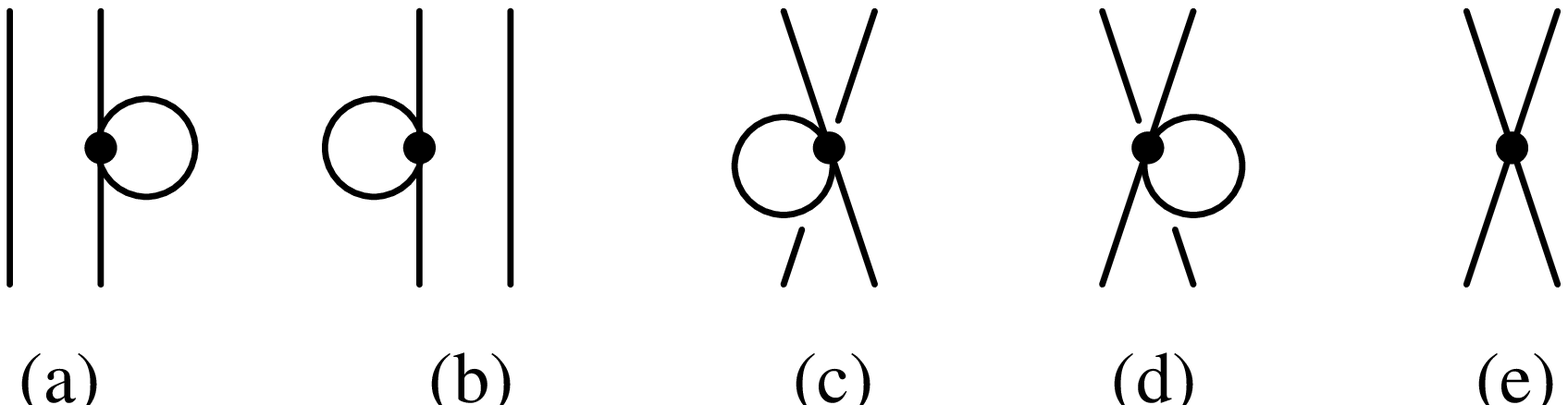}
\vskip0.2cm
\figurecaption{%
Feynman diagrams contributing to the four-point function (5.20).
In this figure time runs vertically from time $0$, 
where the string modes are created, up to time $t$, where they are 
annihilated.
}
\vskip0.0cm
}
\endinsert

\subsection 5.4 Sample calculation

For illustration let us consider the correlation function
\equation{
  \left\langle a^{-}_k(t,p)a^{-}_l(t,q)S_{\rm int}a^{+}_i(0,p)a^{+}_j(0,q)
  \right\rangle_0=
  \left\{M_{kl,ij}t+\cdots\right\}\rme^{-(p+q)t}.
  \enum
}
By a decomposition into 
irreducible $\rmO(d-2)$ tensors,
the interaction matrices $H_{AB}$ 
associated to the states in table~1 with two excited modes
are then easily obtained from the matrix $M_{kl,ij}$.
Here we only discuss the computation of this matrix, which amounts
to the evaluation of the $5$ diagrams shown in fig.~1.

The diagrams (a)--(d) are products of 
the free propagator (5.4) and the 
connected part of the correlation function
\equation{
  \left\langle a^{-}_k(t,p)S_{\rm int}a^{+}_l(0,p)
  \right\rangle_0=
  \left\{M_{kl}t+\cdots\right\}\rme^{-pt}.
  \enum
}
Using Wick's rule the combinatorial factors
are easily obtained from this expression. 
Moreover, from eqs.~(5.2) and (5.3) one infers that 
a non-zero contribution to the diagram is obtained
only when the interaction vertex is inserted at any time
between $0$ and $t$. The part of the Feynman integral that 
adds to the matrix element $M_{kl}$ is then easily identified,
and its value may finally be calculated
following the lines of appendix~A.
In the case of the diagram (a), for example, this leads to 
the result 
\equation{
   M^{(a)}_{kl,ij}=
   \delta_{ki}\delta_{lj}{pq^2\over4}
   \left\{
   \left[\left(d-1\right)qr-{\pi\over6}\right]2c_2+
   \left[\left(d+1\right)qr-{\pi\over6}\left(d-1\right)\right]c_3
   \right\}
   \enum
}
at $D=2$.

To evaluate the tree diagram (e), we only need to substitute the 
free propagators (5.2) and (5.3) for the lines 
from the sources to the interaction
vertex. It again follows from these formulae that 
the non-zero contributions are obtained when
the vertex is at times between $0$ and $t$.
Some algebra then yields
\equation{
  M^{(e)}_{kl,ij}=
  p^2q^2r\,\Bigl\{
  \left[\delta_{ki}\delta_{lj}\delta_{pq}+\delta_{kj}\delta_{li}
  +\delta_{kl}\delta_{ij}\left(1+\delta_{pq}\right)\right]c_2
  \noenum
  \next{2ex}
  \phantom{M^{(e)}_{kl,ij}=\,}
  +\left[\delta_{ki}\delta_{lj}\left(1+\frac{1}{2}\delta_{pq}\right)+
  \delta_{kj}\delta_{li}\left(\frac{1}{2}+\delta_{pq}\right)
  +\frac{1}{2}\delta_{kl}\delta_{ij}\left(1+\delta_{pq}\right)\right]c_3
  \Bigr\}
  \enum
}
for this diagram.

\subsection 5.5 Why is $c_3=-2c_2$ a special case?

If the couplings are related in this way, there are some
important simplifications in the Feynman rules that allow 
the matrix elements $H_{AB}$ to be calculated in closed form
at any level $n$.
We first observe that the interaction at $D=2$ may be 
written in the form
\equation{
  S_{\rm int}=-\frac{1}{4}c_2\int\rmd^2z
  \left(\partial_{+}h\partial_{+}h\right)
  \left(\partial_{-}h\partial_{-}h\right),
  \qquad \partial_{\pm}=\partial_0\pm i\partial_1,
  \enum
}
if $c_3=-2c_2$. In tree diagrams the source-to-vertex propagators
are thus
\equation{
  \left\langle  a^{+}_k(z_0,p)\partial_{\pm}h_l(w)\right\rangle_0
  =\pm
  \delta_{kl}\theta(w_0-z_0)ip\kern1pt\rme^{-p(w_0-z_0)}\rme^{\pm ipw_1},
  \enum
  \next{2ex}
  \left\langle  a^{-}_k(z_0,p)\partial_{\pm}h_l(w)\right\rangle_0
  =\pm
  \delta_{kl}\theta(z_0-w_0)ip\kern1pt\rme^{-p(z_0-w_0)}\rme^{\mp ipw_1},
  \enum
}
and using these expressions it can be shown that the diagrams vanish,
except when the states connected to 
the vertex are as in the diagram
(e) considered above. Moreover, as can be seen from eq.~(5.23), the
contribution is then simply proportional to the
identity matrix.

The only other diagrams are tadpole diagrams of the type (a)--(d),
which are also proportional to the identity matrix. 
Combining all contributions it turns out that the interaction
is diagonalized by the Fock states 
$|n_1,\ldots,n_m;k_1,\ldots,k_m\rangle$
and that the
first-order energy shifts $E_n^1=[E_n^1]$ are given by eq.~(4.11). 
Whether there is a deeper reason for this (a hidden
symmetry perhaps) is, however, not known to us.

\section 6. Summary of results

\vskip-3.5ex

\subsection 6.1 Energy spectrum for general $d$, $c_2$ and $c_3$

As explained above, the obvious symmetries of the effective theory
do not exclude that the $n=3$ vector multiplets mix with each other
once the interactions are switched on.
The detailed calculation
shows, however, that this does not happen to the order of perturbation 
theory considered. For $n\leq3$ the 
first-order energy shifts are then all of the form
\equation{
  E_{n,i}^1=E_{0,1}^1+{\pi^2\over r^3}
  \left\{n\left[\frac{1}{12}\left(d-2\right)-n\right]c_2+
  \nu_{n,i}\left(c_3+2c_2\right)\right\},
  \enum
}
where the energy shift $E_{0,1}^1=[E_0^1]$ of the ground state 
is given by eq.~(4.9)
and the coefficients $\nu_{n,i}$ are listed in table~2%
\kern1pt\footnote{$\dagger$}{\footnotefont%
One readily checks that
the average energy shifts $[E_n^1]$ obtained 
in section~4 from the partition function and
the energy values quoted here satisfy the 
sum rule $\sum_i w_{n,i}E_{n,i}^1=w_n[E_n^1]$.}.

\midinsert
\newdimen\digitwidth
\setbox0=\hbox{\rm 0}
\digitwidth=\wd0
\catcode`@=\active
\def@{\kern\digitwidth}
\vskip-0.6cm
\tablecaption{Values of the coefficients $\nu_{n,i}$ in eq.~(6.1)} 
\vskip1.0ex
$$\vbox{\settabs\+&%
                  xxxxxx&xx&
                  xxxxxxxxxxxx&xxxxxxxx&
                  xxxxxx&xx&
                  xxxxxxxxxxxx&\cr
\thicktablerule
\vskip1.0ex
                \+& \hfill$n,i$\hfill
                 && \hfill$\nu_{n,i}$\hfill
                 && \hfill$n,i$\hfill
                 && \hfill$\nu_{n,i}$\hfill
                 &\cr
\vskip1.0ex
\thintablerule
\vskip1.5ex
  \+& \hfill $1,1$\hfill
  &&  \hskip1ex$\frac{1}{24}\left(5d+7\right)$\hfill 
  &&  \hfill $3,2$\hfill
  &&  \hskip1ex$\frac{1}{8}\left(17d+19\right)$\hfill
  &\cr
\vskip1.0ex
  \+& \hfill $2,1$\hfill
  &&  \hskip1ex$\frac{1}{12}\left(11d+13\right)$\hfill 
  &&  \hfill $3,3$\hfill
  &&  \hskip1ex$\frac{1}{8}\left(9d+43\right)$\hfill
  &\cr
\vskip1.0ex
  \+& \hfill $2,2$\hfill
  &&  \hskip1ex$\frac{1}{12}\left(11d+13\right)$\hfill 
  &&  \hfill $3,4$\hfill
  &&  \hskip1ex$\frac{1}{8}\left(9d+35\right)$\hfill
  &\cr
\vskip1.0ex
  \+& \hfill $2,3$\hfill
  &&  \hskip1ex$\frac{1}{12}\left(5d+25\right)$\hfill 
  &&  \hfill $3,5$\hfill
  &&  \hskip1ex$\frac{1}{8}\left(9d+19\right)$\hfill
  &\cr
\vskip1.0ex
  \+& \hfill $3,1$\hfill
  &&  \hskip1ex$\frac{1}{8}\left(17d+19\right)$\hfill 
  &&  \hfill $3,6$\hfill
  &&  \hskip1ex$\frac{1}{8}\left(5d+43\right)$\hfill
  &\cr
\vskip1.5ex
\thicktablerule
}
$$
\vskip-0.2cm
\endinsert

It follows from these results that 
most degeneracies of the energy levels are lifted
by the interactions if $c_3+2c_2$ does not vanish.
Equation (6.1) also shows that 
the first-order energy shifts tend to increase with $n$
roughly like $n^2$. 
The distances $r$ at which
the large-$r$ expansion of the energy values
applies must therefore be expected to scale with $n$.

\subsection 6.2 Energy spectrum in $d=3$ dimensions

Since $h(z)$ has only one component in $3$ dimensions, the 
interactions $S_2$ and $S_3$ are the same and all energy shifts $E_{n,i}^1$
are therefore proportional to $c_2+c_3$. Without loss we may set
$c_3=-2c_2$, for example, which shows at once that this is a case where the 
free-string degeneracies of the energy levels are not lifted, 
i.e.~to this order of perturbation theory there is
a single energy value $E_n$ for each $n$.

The duality constraint (3.9) now implies $c_2=(2\sigma)^{-1}$
and the expansion
\equation{
  E_n=\sigma r+\mu+{\pi\over r}\left(n-\frac{1}{24}\right)
  -{\pi^2\over2\sigma r^3}\left(n-\frac{1}{24}\right)^2
  +\rmO(r^{-4})
  \enum
}
of the energy values to this order thus involves
no unknown parameters other than the
string tension $\sigma$ and the mass $\mu$. 
Note that the ``multiplets" listed in table~1
are either one-dimensional or do not occur in three dimensions.

\topinsert
\vbox{
\vskip0.0cm
\epsfxsize=5.8cm\hskip2.8cm\epsfbox{rexp.eps}
\vskip0.4cm
\figurecaption{%
String excitation energies $\Delta E_n=E_n-E_0$
in four dimensions [eq.~(6.3)] for coupling
$c_2=(2\sigma)^{-1}$. Physical units 
are defined by the Sommer scale $r_0=0.5\,\fm$
and we have assumed that $r_0^2\sigma=1.36$,
as suggested by recent simulations of the SU(3) gauge theory.
}
\vskip0.1cm
}
\endinsert

\subsection 6.3 Energy spectrum in $d=4$ dimensions

Open--closed string duality requires $c_3=-2c_2$ in this case,
so that there is again only a single energy value $E_n$ for each $n$.
The coupling $c_2$ remains unconstrained,
however,
and the energies
\equation{
  E_n=\sigma r+\mu+{\pi\over r}\left(n-\frac{1}{12}\right)
  -c_2{\pi^2\over r^3}\left(n-\frac{1}{12}\right)^2
  +\rmO(r^{-4})
  \enum
} 
depend on this coupling in a non-trivial way.

For illustration only, we plot the three lowest excitation energies
in fig.~2,
in units of $\pi/r$ and setting $c_2=(2\sigma)^{-1}$.
Taking this value as an example is suggested by 
our results in three dimensions and also
by the fact that the coupling derived from the classical Nambu--Goto action
has this value (the relation $c_3=-2c_2$ holds in this theory too).
The figure shows quite clearly
that the corrections to the free-string energy 
splittings grow with $n$ and that $r$ must therefore be
taken to increasingly large values to reach 
the asymptotic regime.

\section 7. Concluding remarks

The calculations presented in this paper
provide some insight into the effects
of the string self-interactions on the energy spectrum
in the effective string theory. In this approach
the possible form of the interaction terms is constrained
by symmetry considerations only, and the results are
thus representative of a whole class of string theories
rather than of any particular model. 

An important qualitative outcome in space-time dimensions $d\leq4$
is that the free-string degeneracies of the energy levels are 
preserved to first order in the interactions.
Since the number of interaction terms is rapidly increasing,
it seems a bit unlikely, though, that the degeneracies will persist
at higher orders of the perturbation expansion (unless the couplings
assume some special values).

Our calculations also show that the first-order correction to the
ground-state energy (the $1/r^3$ term) is significantly smaller than
the corrections to the excited-state energies. Maybe this
is why string behaviour in the static quark potential in 
${\rm SU}(N)$ gauge theories is seen to set in at surprisingly
short distances [\ref{LWstring}],
where the energy spectrum is otherwise still far from being string-like
[\ref{KutiEtAlI}].
However, the question that will have to be answered first is
whether the low-lying energy values in the gauge theory are
indeed matched by the string-theory formulae (6.2) and (6.3) at 
large distances $r$.
The issue can presumably be resolved
in lattice gauge theory, using numerical simulations,
although this will require a dedicated effort and maybe 
more powerful simulation techniques to be able to 
reach the relevant range of distances with controlled
systematic errors.

\vskip1ex
We are indebted to Hiroshi Suzuki and Wolfgang Lerche for their comments 
on the notion of open--closed string duality, and one of us
(M.L.) wishes to thank the Yukawa Institute at Kyoto,
where part of this paper was written, for the 
generous hospitality extended to him.

\appendix A. Dimensional regularization

The effective string theory lives on a manifold with boundary
where the standard momentum-space techniques of dimensional regula\-ri\-zation
do not apply.
There is a more general formulation 
of dimensional regularization, however,
that can be used here since it 
is based on the heat-kernel representation of the 
propagator in position space
[\ref{MLdimreg}].

\subsection A.1 Heat kernels

The coordinates of a point $z$ on the world-sheet are 
$z_0,z_1,\ldots,z_{D-1}$, where the last $D-2$ parametrize
the unphysical extra dimensions. We take these to be 
a product of circles, with circumference $L$,
and now first need an explicit representation of 
the free massless propagator on this manifold, with 
Dirichlet boundary conditions
at $z_1=0,r$ and periodic boundary conditions in all other 
directions.

To this end we introduce two types of one-dimensional
heat kernels,
the first
\equation{
  K_t^{\rm P}(x,y;T)={1\over T}\sum_{p_0}
  \rme^{ip_0(x-y)}\rme^{-tp_0^2},
  \quad
  p_0={2\pi n_0\over T},\quad n_0\in\gz,
  \enum
}
for periodic boundary conditions and the second
\equation{
  K_t^{\rm D}(x,y;r)={2\over r}\sum_{p_1}
  \sin(p_1x)\sin(p_1y)\rme^{-tp_1^2},
  \quad
  p_1={\pi n_1\over r},\quad n_1\in\nz,
  \enum
}
for Dirichlet boundary conditions.
The heat kernel and the propagator
on the world-sheet  
are then given by
\equation{
  K_t(z,w)=K_t^{\rm P}(z_0,w_0;T)K_t^{\rm D}(z_1,w_1;r)
  \prod_{\alpha=2}^{D-1}K_t^{\rm P}(z_{\alpha},w_{\alpha};L),
  \enum
  \next{1.5ex}
  G(z,w)=\int_0^{\infty}\rmd t\,K_t(z,w).
  \enum
}
Evidently these formulae only make sense for non-negative integral
values of $D-2$, but an analytic continuation in the dimension
of the world-sheet 
is possible later, after the positions
of the vertices in the Feynman diagrams have been integrated over.

Using the Poisson summation formula, the one-dimensional heat kernels
can also be written in the form
\equation{
  K_t^{\rm P}(x,y;T)=
  (4\pi t)^{-1/2}\sum_{n=-\infty}^{\infty}
  \rme^{-(x-y+nT)^2/4t},
  \enum
  \next{1.5ex}
  K_t^{\rm D}(x,y;r)=
  (4\pi t)^{-1/2}\sum_{n=-\infty}^{\infty}
  \left\{\rme^{-(x-y+2nr)^2/4t}-\rme^{-(x+y+2nr)^2/4t}\right\}.
  \enum
}
In particular, in the limit $t\to0$ the sums
\equation{
  \kappa_0(t)=\sum_{p_0}\rme^{-tp_0^2},
  \qquad
  \kappa_1(t)=\sum_{p_1}\rme^{-tp_1^2}
  \enum
}
(where $p_0$ and $p_1$ run over the values specified above)
are equal to
$T(4\pi t)^{-1/2}$ and $-1/2+r(4\pi t)^{-1/2}$ respectively, up
to exponentially small terms.

\subsection A.2 Calculation of $\left\langle S_1\right\rangle_0$

Since the final results do not depend on the 
compactification scale $L$, we may now just as well set $L=T$ 
in this subsection.
When the heat-kernel representation (A.4)
of the propagator is inserted in eq.~(2.13), the integral
\equation{
  \left\langle S_1\right\rangle_0=
  -{b\over r}\left(d-2\right)
  \int_0^{\infty}\rmd t\,\left[\kappa_0(t)\right]^{D-1}
  {\rmd\over\rmd t}\kappa_1(t)
  \enum
}
is obtained, which is absolutely convergent
for all complex $D$ in the range $\Re D<0$. Moreover, 
by adding and subtracting the asymptotic term $T(4\pi t)^{-1/2}$, raised to the
power $D-1$,
from the first factor in the integrand, the expression can be
analytically continued in $D$ to the whole complex plane.
This leads to the representation
\equation{
  \left\langle S_1\right\rangle_0=
  -{b\over r}\left(d-2\right)
  \left\{{\pi T\over24r}+\int_0^{\infty}\rmd t
  \left[\kappa_0(t)-T(4\pi t)^{-1/2}\right]
  {\rmd\over\rmd t}\kappa_1(t)
  \right\}.
  \enum
}
at $D=2$.

We may now expand $\kappa_0(t)$ 
and $\kappa_1(t)$
in the exponentials $\rme^{-(mT)^2/4t}$ and $\rme^{-(n\pi/r)^2t}$
respectively and integrate the terms of the double-series 
over the parameter $t$. The sum over $m$
can then be performed analytically, and the formula
\equation{
  \left\langle S_1\right\rangle_0=
  -b\left(d-2\right){\pi T\over r^2}\,\biggl\{{1\over24}-\sum_{n=1}^{\infty}
  {nq^n\over1-q^n}\biggr\}
  \enum
}
is thus obtained,
which coincides with eq.~(2.14).

\subsection A.3 More complicated diagrams

The one-particle irreducible parts of the Feynman diagrams 
considered in this paper are tadpole diagrams where one or more
loops are attached to a single vertex.
All these diagrams can be worked
out analytically along the lines explained above.

In the first step the heat-kernel representation (A.4) 
is substituted for all (internal) propagators and the integral
over the world-sheet coordinates of the interaction vertex is 
performed. One is then left with an integral over
the heat-kernel parameters that is absolutely convergent 
for sufficiently small $\Re D$. 
The analytic continuation 
to the physical point $D=2$ is finally achieved by
separating the leading singularities of the integrand
at small values of the  heat-kernel parameters.

\appendix B. Radon transform of spherically symmetric functions

In this appendix we derive the expansion (3.2) from the 
spectral representation (3.1) and the fact that
the Polyakov loop correlation function is spherically symmetric.
The key point to note is that the left-hand side of eq.~(3.1)
has the form of an integral transformation that
maps the correlation function to some function of 
$z=|x_1-y_1|$. This is actually a special case of the 
Radon transform [\ref{Gelfand}], and we now first
discuss some general properties of this transformation.

\subsection B.1 Definition

In the following we consider smooth functions $f(r)$ on the 
half-line $r>0$
that fall off faster than any inverse power of $r$ at infinity.
Their behaviour in the limit $r\to0$ is left unspecified and 
can be arbitrarily singular. 

For any integer $n\geq2$ and all $z>0$, we define the 
Radon transform $\R_nf(z)$
of a given function $f(r)$ through
\equation{
  \R_n f(z)=\int_{\rz^{n-1}}\rmd^{n-1}y
  \left.f(r)\right|_{r=\sqrt{y^2+z^2}}.
  \enum
}
If we interpret $r$ as the radius in $\rz^n$, the integral on the right
of this equation is taken over an $n-1$ dimensional 
hyperplane, parametrized by the coordinates
$y\in\rz^{n-1}$, at a distance $z$ from the origin.
It is easy to show that 
the transformed function is again in the class of functions
specified above, and the transformation can thus be applied repeatedly.
An immediate consequence of the definition (B.1) is then that 
\equation{
  \R_n\R_m=\R_{n+m-1}.
  \enum
}
In particular, $\R_n=\R_2\R_2\ldots \R_2$ ($n-1$ factors).

\subsection B.2 Inversion of the Radon transform

In $n=3$ dimensions we have
\equation{
  \R_3f(z)=2\pi\int_{z}^{\infty}\rmd r\,rf(r),
  \enum
}
and the function $f(r)$ can be reconstructed from its
Radon transform through
\equation{
  f(r)=-(2\pi r)^{-1}{\rmd\over\rmd r}\R_3f(r).
  \enum
}
The Radon transform $\R_3$ is thus an invertible mapping from the space
of functions defined above onto itself. 

In view of the semi-group property (B.2), the same is also true for all
other trans\-formations. In particular, the inverse of $\R_2$
is obtained by first acting with $\R_2$ and then using eq.~(B.4).
Explicitly we have
\equation{
  \R_n^{-1}=
  \cases{(\R_3^{-1})^{{1\over2}(n-1)} & if $n$ is odd, \cr
         \noalign{\vskip2ex}
         (\R_3^{-1})^{{1\over2} n}\R_2 & if $n$ is even,\cr}
  \enum
}
where $\R_3^{-1}f(r)=-(2\pi r)^{-1}f'(r)$.

\subsection B.3 Example

It is now almost trivial to find the function $k(r)$ that satisfies
\equation{
  \R_nk(z)=\rme^{-Ez}.
  \enum
}
If $n$ is odd, say $n=2\nu+1$, it is given by
\equation{ 
  k(r)=
  \left(-2\pi\right)^{-\nu}
  \left({1\over r}{\rmd\over\rmd r}\right)^{\nu}\rme^{-Er}
  =
  2r\left(E\over2\pi r\right)^{n\over2}
  K_{{n\over2}-1}(Er),
  \enum
}
where the normalization of the Bessel functions is as in 
ref.~[\ref{GR}].
In the other case the solution is obtained by noting that
\equation{
  \R_2f(z)=2zK_1(Ez)\quad\hbox{if}\quad f(r)=\rme^{-Er}.
  \enum
}
The application of eq.~(B.5) for $n=2\nu$ then yields
\equation{
  k(r)=
  \left(-2\pi\right)^{-\nu}
  \left({1\over r}{\rmd\over\rmd r}\right)^{\nu}
  2rK_1(Er)
  =
  2r\left(E\over2\pi r\right)^{n\over2}
  K_{{n\over2}-1}(Er),
  \enum
}
i.e.~the same expression
in terms of the Bessel functions as above.

\subsection B.4 Proof of eq.~(3.2)

As already mentioned at the beginning of this appendix,
eq.~(3.1) may be written in the form
\equation{
  \R_{d-1}f(z)=\sum_{n=0}^{\infty}\left|v_n\right|^2\rme^{-\tilde{E}_nz},
  \enum
}
where $f(r)$ now stands for 
the Polyakov loop correlation function at distance $r$.
The application of the inverse Radon transform and the
results obtained in the previous subsection
then immediately lead to the expansion (3.2). 

A technical point we would like to emphasize here is that the inverse
of the Radon transform is unique and that this crucial property holds
irrespective of the behaviour of the functions at small distances $r$. 
The Radon
transform is actually local in the sense that the values of the
transformed function at $z\geq a$ only depend on the values of the
input function at distances $r\geq a$, for any fixed $a>0$.

\appendix C. Open--closed string duality and the Nambu--Goto string

The energy spectrum of a Nambu--Goto string with fixed ends,
\equation{
  E_n=\sigma r\left\{1+{2\pi\over\sigma r^2}
  \left[-\frac{1}{24}\left(d-2\right)+n\right]
  \right\}^{1/2},
  \qquad n=0,1,2,\ldots,
  \enum
}
was calculated
some time ago by Arvis [\ref{Arvis}]
through formal canonical quantization.
Since the degeneracies of the energy eigenstates
are the same as in the case of the free string, the 
associated partition function reads
\equation{
  \Z=\sum_{n=0}^{\infty}w_n\rme^{-E_nT}.
  \enum
}
We now derive the 
closed-string representation of this partition function.

Our starting point is the formula
\equation{
  \Z={1\over\sqrt{4\pi}}\int_0^{\infty}{\rmd t\over t^{3/2}}\,
  \rme^{-{1\over 4t}-(E_0T)^2t}\,
  \Bigl\{q^{-{1\over24}}\eta(q)\Bigr\}^{2-d},
  \qquad
  q=\rme^{-2\pi\sigma T^2t},
  \enum
}
which is easily verified by expanding 
the curly bracket in powers of $q$
and by integrating the resulting series term by term.
The modular transformation property of the $\eta$-function
may now be used and this leads, after re-expanding the curly bracket,
to the representation
\equation{
  \Z={(\sigma T^2)^{{1\over2}(d-2)}\over\sqrt{4\pi}}
  \int_0^{\infty}\rmd t\, t^{{1\over2}(d-5)}\,
  \sum_{n=0}^{\infty}w_n\exp\biggl\{
  -{\tilde{E}_n^2\over 4(\sigma T)^2t}-(\sigma Tr)^2t\biggr\}
  \enum
}
in which the closed-string energies are given by
\equation{
  \tilde{E}_n=\sigma T
  \left\{1+{8\pi\over\sigma T^2}\left[-\frac{1}{24}\left(d-2\right)+n\right]
  \right\}^{1/2}.
  \enum
}
Integration of the terms in the sum (C.4) finally yields 
a series of Bessel functions of the form (3.2), with
\equation{
   |v_n|^2={\sigma T\over\tilde{E}_n}
   \left({\pi\over\sigma}\right)^{{1\over2}(d-2)}w_n,
   \enum
}
and open--closed string duality thus holds exactly.

\beginbibliography


\bibitem{LWstring}
M. L\"uscher, P. Weisz,
J. High Energy Phys. 07 (2002) 049

\bibitem{Pushan}
P. Majumdar,
Nucl. Phys. B664 (2003) 213


\bibitem{CaselleEtAl}
M. Caselle, R. Fiore, F. Gliozzi, M. Hasenbusch, P. Provero,
Nucl. Phys. B486 (1997) 245

\bibitem{CasellePaneroProvero}
M. Caselle, M. Panero, P. Provero,
J. High Energy Phys. 0206 (2002) 061

\bibitem{CaselleHasenbuschPaneroI}
M. Caselle, M. Hasenbusch, M. Panero,
J. High Energy Phys. 0301 (2003) 057;
Nucl. Phys. (Proc. Suppl.) 129-130 (2004) 593

\bibitem{CasellePepeRago}
M. Caselle, M. Pepe, A. Rago,
Nucl. Phys. (Proc. Suppl.) 129-130 (2004) 721;
hep-lat/0406008

\bibitem{CaselleHasenbuschPaneroII}
M. Caselle, M. Hasenbusch, M. Panero,
hep-lat/0403004


\bibitem{MichaelPerantonis}
S. Perantonis, C. Michael,
Nucl. Phys. B347 (1990) 854

\bibitem{KutiEtAlI}
K. J. Juge, J. Kuti, C. J. Morningstar,
Nucl. Phys. B (Proc.Suppl.) 63 (1998) 326;
{\it ibid.}\/ 73 (1999) 590;
{\it ibid.}\/ 83 (2000) 503;
{\it ibid.}\/ 106 (2002) 691;
Phys. Rev. Lett. 90 (2003) 161601


\bibitem{ParisiPetronzioRapuano}
G. Parisi, R. Petronzio, F. Rapuano,
Phys. Lett. B128 (1983) 418

\bibitem{DeForcrandEtAl}
P. de Forcrand, G. Schierholz, H. Schneider, M. Teper,
Phys. Lett. B160 (1985) 137

\bibitem{OhtaFukugitaUkawa}
S. Ohta, M. Fukugita, A. Ukawa,
Phys. Lett. B173 (1986) 15


\bibitem{LuciniTeper}
B. Lucini, M. Teper,
Phys. Rev. D64 (2001) 105019

\bibitem{KutiEtAlII}
K. J. Juge, J. Kuti, F. Maresca, C. Morningstar, M. J. Peardon,
Nucl. Phys. (Proc. Suppl.) 129-130 (2004) 703


\bibitem{Nambu}
Y. Nambu,
Phys. Lett. B80 (1979) 372

\bibitem{WKB}
M. L\"uscher, K. Symanzik, P. Weisz,
Nucl. Phys. B173 (1980) 365

\bibitem{UniversalTerm}
M. L\"uscher,
Nucl. Phys. B180 (1981) 317


\bibitem{QFTbI}
K. Symanzik,
Nucl. Phys. B190 [FS3] (1981) 1

\bibitem{QFTbII}
M. L\"uscher,
Nucl. Phys. B254 (1985) 52


\bibitem{DietzFilk}
K. Dietz, T. Filk,
Phys. Rev. D27 (1983) 2944

\bibitem{AmbjornEtAl}
J. Ambj{\o}rn, P. Olesen, C. Peterson,
Phys. Lett. B142 (1984) 410; Nucl. Phys. B244 (1984) 262


\bibitem{TransferI}
M. L\"uscher,
Commun. Math. Phys. 54 (1977) 283

\bibitem{TransferII}
K. Osterwalder, E. Seiler,
Ann. Phys. (NY) 110 (1978) 440

\bibitem{TransferIII}
E. Seiler,
Gauge theories as a problem of constructive quantum field
theory and statistical mechanics, Lecture Notes in Physics 159
(Springer, Berlin, 1982)

\bibitem{TransferIV}
M. L\"uscher,
Selected topics in lattice field theory,
Lectures given at Les Houches (1988),
in: Fields, strings and
critical phenomena, eds. E. Br\'ezin, J. Zinn-Justin
(North-Holland, Amsterdam, 1989)


\bibitem{MLdimreg}
M. L\"uscher,
Ann. Phys. 142 (1982) 359


\bibitem{Gelfand}
I. M. Gelfand, M. I. Graev, N. Y. Vilenkin,
Generalized functions, Vol. 5
(Academic Press, New York, 1966)


\bibitem{GR}
I. S. Gradshteyn, I. M. Ryzhik,
Table of integrals, series and products
(Academic Press, New York, 1965)


\bibitem{Arvis}
J. F. Arvis,
Phys. Lett. B127 (1983) 106

\endbibliography

\bye